\documentclass[
 aip,
 jcp,
 amsmath,
 amssymb,
]{revtex4-1}
\usepackage[utf8]{inputenc}
\usepackage{array}
\usepackage{amssymb}
\usepackage{amsmath}
\usepackage{float}
\usepackage{xcolor}
\usepackage{graphicx}
\usepackage{braket}

\begin{document}

\title{Redundant parameter dependencies in truncated classic and quantum Linear Response and Equation of Motion theory.}
\author{Erik Rosendahl Kjellgren}
\email{kjellgren@sdu.dk}
\affiliation{Department of Physics, Chemistry and Pharmacy,
University of Southern Denmark, Campusvej 55, 5230 Odense, Denmark.}
\author{Peter Reinholdt}
\affiliation{Department of Physics, Chemistry and Pharmacy,
University of Southern Denmark, Campusvej 55, 5230 Odense, Denmark.}
\author{Karl Michael Ziems}
\affiliation{School of Chemistry, University of Southampton, Highfield, Southampton SO17 1BJ, United Kingdom}
\affiliation{Department of Chemistry, Technical University of Denmark, Kemitorvet Building 207, DK-2800 Kongens Lyngby, Denmark.}
\author{Stephan P. A. Sauer}
\affiliation{Department of Chemistry, University of Copenhagen, DK-2100 Copenhagen \O.}
\author{Sonia Coriani}
\affiliation{Department of Chemistry, Technical University of Denmark, Kemitorvet Building 207, DK-2800 Kongens Lyngby, Denmark.}
\author{Jacob Kongsted}
\affiliation{Department of Physics, Chemistry and Pharmacy,
University of Southern Denmark, Campusvej 55, 5230 Odense, Denmark.}
\date{\today}

\begin{abstract}
Extracting molecular properties from a wave function can be done through the linear response (LR) formalism or, equivalently, the equation of motion (EOM) formalism.
For a simple model system, He in a 6-31G basis, it is here shown that calculated excitation energies depend on the specifically chosen orbitals, even when the ground-state is the FCI solution, if the LR is truncated to a singles expansion.
This holds for naive, projected, self-consistent, and state-transfer parametrizations of the LR operators.
With a focus on the state-transfer parameterization, this problem is shown to also hold for more complicated systems, and is also present when the LR is truncated to singles and doubles.
This problem can be alleviated by performing a ground-state constrained trace optimization of the Hessian matrix before performing the LR calculation.
It is finally shown that spectra can be further improved for small LR expansions by targeting only a few states in the constrained trace optimization using constrained state-averaged UCC.
\end{abstract}


\maketitle

\section{Introduction}

One of the main goals of computational chemistry in general, and quantum chemistry in particular, is the ability to predict the properties of molecular systems.
The primary property of interest is often the ground state energy of a molecule and, with it, its corresponding wave function. 
Although important for determining the structure of a molecule or reaction energies, the ground state energy is experimentally not accessible and thus is of less application than properties related to electronic excited states.
Modeling excited states is crucial for all types of electronic spectroscopy, as well as in photochemistry, photophysics, and other diverse research fields such as vision, solar energy harvesting, photodynamic therapy, photosensitizers, photocatalysis, photosynthesis, or radiation damage, to name just a few.
However, the development of electronic structure methods for excited states is more challenging and, arguably, still behind those for ground states, as it faces unique challenges: ($i$) a balanced treatment of multiple states that may have significantly differing electronic characters is needed; ($ii$) the complexity intensifies when the system moves away from the equilibrium geometry of the ground state.
The ability to predict or explain these optical properties of molecules is, therefore, one of the most important applications of computational chemistry in diverse research fields. 

Approaches to obtain electronic excitation energies can be broadly classified into two groups: state-specific (and state-averaged) methods, and response theory/polarization propagator-based methods. 
Within the first category, electronic excitation energies are obtained as the difference between the total energies of individually optimized ground- and excited-state wave functions. 
Within propagator-based formalisms, only the ground-state wave function is explicitly optimized,
whereas the excitation energies are obtained by solving a set of homogeneous linear response equations. 
Arguably, the equation of motion (EOM) approach\cite{Rowe1968, McCurdy1977} or the linear response (LR) approach,\cite{Olsen1985} both developed in combination with different types of wave functions, are nowadays the most used approaches to determine spectroscopic properties across different frequency regions.
Examples of such calculations on classical high-performance computers are plentiful, see e.g., Refs.\citenum{Sauer2009,Silva2010,Veril2021,Fransson:2021}.

With the advent of quantum computing, either in the form of today's noisy intermediate-scale quantum (NISQ) computing or future fault-tolerant quantum computing, the interest in the details of the formulation of both methods has been rekindled.
A quantum chemical calculation on NISQ devices typically starts with the optimization of the wavefunction in the form of 
a variational quantum eigensolver\cite{Peruzzo2014-rx} (VQE) calculation, which nowadays can even be performed in the presence of an environment.\cite{Kjellgren2023,Castaldo2022-aw,Rossmannek2023-rn,Hohenstein2023-qz}
The calculation of molecular ground state properties could then proceed with the calculation of an expectation value with the optimized wave function in order to obtain, e.g., electric field gradients\cite{Nagy2024} or hyperfine coupling constants.\cite{Jensen2025}
A promising route to extract molecular properties beyond the ground state is via the aforementioned LR and EOM formalisms.
When the EOM approach was first introduced in the context of quantum computing it was dubbed quantum EOM (qEOM),\cite{McClean2017, Ollitrault2020} and correspondingly the linear response approach is in the quantum computing regime known as quantum linear response (qLR).\cite{Kumar2023}

Since then, a substantial amount of effort has been put into developing the qEOM and qLR formalisms.
Some of these further developments include the multi-component-EOM approach (mcEOM),\cite{Pavoevi2021a} which allows to treat both electrons and nuclei without invoking the Born-Oppenheimer approximation; the quantum electrodynamics EOM approach (QED-EOM)\cite{Pavoevi2021} for strongly light-matter coupled systems; an extension of the VQE and qEOM algorithms to the band structures of solids;\cite{Fan2021-mb} a combination of the spin-flip formalism with the qEOM approach;\cite{Pavoevi2023} a version of qEOM with self-consistent excitation operators (q-sc-EOM)\cite{Asthana2023} that fulfill the killer condition, and its combination with the Davidson algorithm for solving large eigenvalue problems;\cite{Kim2023} an application of qEOM to the calculation of thermal averages of quantum states;\cite{Morrone2024-yd} a qEOM implementation which includes not only the usual singles and doubles excitations but also selected triples excitations;\cite{Zheng2024-kt} a qEOM approach based on an orbital-optimized variational quantum eigensolver, called the oo-qEOM approach,\cite{Jensen2024-hs} and, finally, an EOM version of the internally contracted multireference unitary coupled-cluster framework (EOM-ic-MRUCC).\cite{Li2025-qc}
An interest in the performance of qEOM compared to other methods of gaining excited state information has also been active with comparison of sc-qEOM and MC-VQE,\cite{Grimsley2024-qo} comparison of naive-qEOM and QSE,\cite{Gandon2024-lo} and comparison of sc-qEOM and QSE.\cite{Kwao2025-ya}

Similarly, the qLR method has recently been thoroughly studied and extended. 
First of all, Ziems et al. \cite{Ziems2024-ar} derived and investigated orbital-optimized qLR (oo-qLR) equations in active spaces for eight different types of parametrizations using naive, self-consistent, state-transfer, and projected excitation operators. 
Reinholdt et al. combined the self-consistent qLR (sc-qLR) approach first with the Davidson algorithm for solving the eigenvalue problem\cite{Reinholdt2024-nl} and then with the polarizable embedding environment in the PE-sc-qLR approach,\cite{Reinholdt2025-dn} while von Buchwald et al. \cite{Von_Buchwald2024-ae} presented a reduced density matrix driven version of the naive oo-qLR approach. 
Finally, Ziems et al.\cite{Ziems2025-tt} presented for the first time UV/Vis spectra calculated with the naive, projected, and all-projected oo-qLR approaches on IBM's quantum devices using several advanced error mitigation techniques.

At the same time, it was observed that parameterizations of EOM or LR that introduce a non-identity metric can lead to poorly conditioned generalized eigenvalue problems.\cite{Kjellgren2024-pc}
Similarly, the problems of a non-identity metric were shown to also cause problems in methods that are very similar to the EOM/LR methods, such as quantum subspace expansion (QSE).\cite{Gandon2024-lo,Kwao2025-ya}

This work investigates how excitation energies obtained within certain LR parametrizations depend on orbital rotation parameters that are redundant with respect to the ground-state energy.
This occurs when the wave function is over-parametrized compared to the excitation level used for the LR part. 
Constrained state-averaged (CSA) UCC is used to circumvent the problem. Alternatively, the same level of theory is used for the ground-state and the LR excitation operators.

\section{Theory}

\subsection{Unitary Coupled Cluster}

The unitary coupled cluster (UCC) wave function is given as
\begin{equation}
    \left|\text{UCC}(\boldsymbol{\theta})\right> = \exp\left(\sum_I \theta_I\left(\hat{T}_I - \hat{T}_I^\dagger\right)\right)\left|\text{HF}\right> 
    \equiv \exp\left(\sum_I\theta_I\hat{\sigma}_I\right)\left|\text{HF}\right>~,
    \label{eq.1}
\end{equation}
the single and double excitation cluster operators being defined in a spin-adapted approach,\cite{Paldus1977,Piecuch1989,Packer1996}
\begin{align}
    \hat{T}_{pq} &= \frac{1}{\sqrt{2}}\hat{E}_{pq}\\
    \hat{T}_{pqrs} &= \frac{1}{2\sqrt{\left(1+\delta_{pr}\right)\left(1+\delta_{qs}\right)}}\left(\hat{E}_{pq}\hat{E}_{rs} + \hat{E}_{ps}\hat{E}_{rq}\right)\\
    \hat{T}^{\prime}_{pqrs} &=\frac{1}{2\sqrt{3}}\left(\hat{E}_{pq}\hat{E}_{rs} - \hat{E}_{ps}\hat{E}_{rq}\right)~,
\end{align}
where $\hat{E}_{pq} = \hat{a}_{p,\alpha}^\dagger\hat{a}_{q,\alpha} + \hat{a}_{p,\beta}^\dagger\hat{a}_{q,\beta}$ is the singlet one-electron excitation operator.
The higher-order cluster operators are defined as
\begin{align}
    \hat{T}_{PQRSTU} &= \hat{a}^\dagger_P\hat{a}^\dagger_R\hat{a}^\dagger_T\hat{a}_U\hat{a}_S\hat{a}_Q\\
    \hat{T}_{PQRSTUMN} &= \hat{a}^\dagger_P\hat{a}^\dagger_R\hat{a}^\dagger_T\hat{a}^\dagger_M\hat{a}_N\hat{a}_U\hat{a}_S\hat{a}_Q\\\
    & \cdots
\end{align}
with the restriction that the number of $\alpha$ creation operators is equal to the number of $\alpha$ annihilation operators in every cluster operator (capital indices indicate spin orbitals).
This, in turn, imposes the same restriction on $\beta$.
This restriction ensures that the operators conserve the number of $\alpha$-electrons and $\beta$-electrons, $N_\alpha$ and $N_\beta$.

The ground state of the UCC wave function can be found by performing a variational minimization of the energy,
\begin{equation}
    E_\text{gs} = \min_{\boldsymbol{\theta}} \left<\text{UCC}(\boldsymbol{\theta})\left|\hat{H}\right|\text{UCC}(\boldsymbol{\theta})\right>
\end{equation}
where $\hat{H}$ is the molecular electronic Hamiltonian,
\begin{equation}
    \hat{H} = \sum_{pq}h_{pq}\hat{E}_{pq} + \frac{1}{2}\sum_{pqrs}g_{pqrs}\hat{e}_{pqrs}~.
\end{equation}
Here, $h_{pq}$ are the one-electron integrals in the molecular orbital (MO) basis, and $g_{pqrs}$ are the two-electron integrals in the MO basis; $\hat{e}_{pqrs} = \hat{E}_{pq}\hat{E}_{rs} - \delta_{qr}\hat{E}_{ps}$ is the singlet two-electron excitation operator.

The unitary coupled cluster wave function can be extended to an orbital-optimized form by including the orbital rotation parameterization,
\begin{equation}
    \left|\text{oo-UCC}(\boldsymbol{\kappa},\boldsymbol{\theta})\right> = \exp\left(\sum_{p>q}\kappa_{pq}\hat{\kappa}_{pq}\right)\left|\text{UCC}(\boldsymbol{\theta})\right>
\end{equation}
with $\hat{\kappa}_{pq}=\frac{1}{\sqrt{2}}\left(\hat{E}_{pq}-\hat{E}_{qp}\right) = \hat{\sigma}_{pq}$.
The orbital rotation parameterization of the wave function is equivalent to performing a unitary transformation of the one- and two-electron integrals,\cite{Helgaker2013-xk}
\begin{align}
    h_{pq}\left(\boldsymbol{\kappa}\right) &= \sum_{p'q'} \left[\mathrm{e}^{\boldsymbol{\kappa}}\right]_{q'q}h_{p'q'}\left[\mathrm{e}^{-\boldsymbol{\kappa}}\right]_{p'p}\label{eq:int1e_kappa},\\
    g_{pqrs}\left(\boldsymbol{\kappa}\right) &= \sum_{p'q'r's'}\left[\mathrm{e}^{\boldsymbol{\kappa}}\right]_{s's}\left[\mathrm{e}^{\boldsymbol{\kappa}}\right]_{q'q}g_{p'q'r's'}\left[\mathrm{e}^{-\boldsymbol{\kappa}}\right]_{p'p}\left[\mathrm{e}^{-\boldsymbol{\kappa}}\right]_{r'r}\label{eq:int2e_kappa}~,
\end{align}
thus leading to a new minimization problem to obtain the ground-state energy,
\begin{equation}
    E_\text{gs} = \min_{\boldsymbol{\theta},\boldsymbol{\kappa}} \left<\text{UCC}(\boldsymbol{\theta})\left|\hat{H}(\boldsymbol{\kappa})\right|\text{UCC}(\boldsymbol{\theta})\right>~.
\end{equation}
With the introduction of orbital rotation parameters, redundant parameters may occur, as the UCC parameterization and orbital rotation parameterization might overlap.
An orbital rotation parameter is identified as redundant if
\begin{equation}
    \min_{\boldsymbol{\theta},\boldsymbol{\kappa}} \left<\text{UCC}(\boldsymbol{\theta})\left|\hat{H}(\boldsymbol{\kappa})\right|\text{UCC}(\boldsymbol{\theta})\right> = \left.\min_{\boldsymbol{\theta},\boldsymbol{\kappa}\setminus\kappa_{pq}} \left<\text{UCC}(\boldsymbol{\theta})\left|\hat{H}(\boldsymbol{\kappa})\right|\text{UCC}(\boldsymbol{\theta})\right>\right|_{\kappa_{pq}\in \mathbb{R}}
    \label{eq:kappa_red_def}
\end{equation}
that is, an orbital rotation parameter is redundant if the ground-state energy can be recovered for all possible values of the orbital rotation parameter.
Similarly, redundant UCC amplitudes can be defined as
\begin{equation}
 \min_{\boldsymbol{\theta},\boldsymbol{\kappa}} \left<\text{UCC}(\boldsymbol{\theta})\left|\hat{H}(\boldsymbol{\kappa})\right|\text{UCC}(\boldsymbol{\theta})\right> = \left.\min_{\boldsymbol{\theta}\setminus\theta_{i},\boldsymbol{\kappa}} \left<\text{UCC}(\boldsymbol{\theta})\left|\hat{H}(\boldsymbol{\kappa})\right|\text{UCC}(\boldsymbol{\theta})\right>\right|_{\theta_{i}\in \mathbb{R}}
    \label{eq:theta_red_def}
\end{equation}
These definitions of redundant parameters are general and apply to other parameterizations of the wave function.

\subsection{Linear Response}

In this work, we explicitly consider the LR formalism instead of the EOM formalism.
However, we note that for a unitary coupled cluster wave function, LR and EOM are identical.\cite{Taube2006}
Therefore, conclusions made about LR are also valid for EOM.
In this section, we will only present the working equations for the calculation of excitation energies using LR.
We refer the reader to J{\o}rgensen and Olsen \cite{Olsen1985} for a detailed derivation of LR, and to
Ziems et al.\cite{Ziems2024-ar}
for different parameterizations using unitary coupled cluster.

For the calculation of excitation energies using LR the following (generalized) eigenvalue problem must be solved
\begin{align}
 \left( \boldsymbol{E}^{[2]} - \varepsilon_i \boldsymbol{S}^{[2]} \right)\boldsymbol{V}_i = \boldsymbol{0}\label{eq:lr},
\end{align}
where $\varepsilon_i$ is the $i$'th excitation energy (eigenvalue) and $\boldsymbol{V}_i$ is the corresponding excitation vector (eigenvector).
It should be noted that the eigenvalues come in pairs of $+|\varepsilon_i|$ and $-|\varepsilon_i|$, for the rest of this work, when referring to the eigenvalues or excitation energies, we will only be referring to the positive subset. 
The Hessian matrix ($\boldsymbol{E}^{[2]}$) and the metric matrix ($\boldsymbol{S}^{[2]}$) are defined as
\begin{align}
 \boldsymbol{E}^{[2]} &= \begin{pmatrix}
    \boldsymbol{A} & \boldsymbol{B}\\
      \boldsymbol{B}^* & \boldsymbol{A}^*           
     \end{pmatrix}, \quad 
     \boldsymbol{S}^{[2]} = \begin{pmatrix}
    \boldsymbol{\Sigma} & \boldsymbol{\Delta}\\
     -\boldsymbol{\Delta} ^* &  -\boldsymbol{\Sigma}^*           
     \end{pmatrix}~. \quad
\end{align}
The submatrices of the Hessian matrix and metric matrix are defined as
\begin{align}
    \boldsymbol{A} &= \boldsymbol{A}^\dagger,\quad A_{IJ} =  \left<0\left|\left[\hat{R}_{I}^\dagger,\hat{H},\hat{R}_{J}\right]\right|0\right>\\
    \boldsymbol{B} &= \boldsymbol{B}^\mathrm{T},\quad B_{IJ} =  \left<0\left|\left[\hat{R}_{I}^\dagger,\hat{H},\hat{R}_{J}^\dagger\right]\right|0\right>\\
    \boldsymbol{\Sigma} &= \boldsymbol{\Sigma}^\dagger,\quad \Sigma_{IJ} = \left<0\left|\left[\hat{R}_{I}^\dagger,\hat{R}_{J}\right]\right|0\right>\label{eq:Sigma}\\
    \boldsymbol{\Delta} &= -\boldsymbol{\Delta}^\mathrm{T},\quad \Delta_{IJ} = \left<0\left|\left[\hat{R}_{I}^\dagger,\hat{R}_{J}^\dagger\right]\right|0\right>.
\end{align}
with $\left|0\right>=\left|\text{UCC}(\boldsymbol{\theta})\right>$ being the ground-state UCC reference wave function, Eq. (\ref{eq.1}), and $\hat{R}$ being an excitation operator.
The specific form of $\hat{R}$ depends on the parametrization used.
To specify $\hat{R}$, let us first introduce some base excitation operators,
\begin{align}
    \hat{G}_{ai} &= \frac{1}{\sqrt{2}}\hat{E}_{ai}\label{eq:G1}\\
    \hat{G}_{aibj} &= \frac{1}{2\sqrt{\left(1+\delta_{ab}\right)\left(1+\delta_{ij}\right)}}\left(\hat{E}_{ai}\hat{E}_{bj} + \hat{E}_{aj}\hat{E}_{bi}\right)\label{eq:G2}\\
    \hat{G}^{\prime}_{aibj} &= \frac{1}{2\sqrt{3}}\left(\hat{E}_{ai}\hat{E}_{bj} - \hat{E}_{aj}\hat{E}_{bi}\right).
\end{align}
In the above definitions, the indices $i$ and $j$ refer to occupied orbitals in the Hartree-Fock reference, and $a$ and $b$ refer to unoccupied orbitals in the Hartree-Fock reference.
These base excitation operators are the singlet single excitation operators and singlet double excitation operators, thus guaranteeing that only singlet excitations will be found when solving the LR equation.
Limiting the LR equations to singles and doubles will be referred to as LRSD.
The different types of parametrizations are the following
\begin{align}
    &\hat{R}_I^\text{naive} = \hat{G}_I\\
    &\hat{R}_I^\text{proj} = \hat{G}_I\left|0\right>\left<0\right| - \left<0\left|\hat{G}_I\right|0\right>\\
    &\hat{R}_I^\text{sc} = \boldsymbol{U}\hat{G}_I\boldsymbol{U}^\dagger\\
    &\hat{R}_I^\text{st} = \boldsymbol{U}\hat{G}_I\left|\text{HF}\right>\left<0\right|
\end{align}
with $I$ being used as a compound index for $ai$ and $aibj$.

Since the state-transfer and self-consistent parametrizations will have special significance in the following sections, their working equations will be shown here.
For the working equations of the naive and projected parametrizations, we refer to the overview given by Ziems et al.\cite{Ziems2024-ar}
The explicit matrix elements of the state-transfer parametrization are
\begin{align}
    A_{IJ}^{[2],\text{st}} &= \left<\text{HF}\left|\hat{G}_I^\dagger\boldsymbol{U}^\dagger\hat{H}\boldsymbol{U}\hat{G}_J\right|\text{HF}\right> - \delta_{IJ}E_0\\ B_{IJ}^{[2],\text{st}} &= 0\\
    \Sigma_{IJ}^{[2],\text{st}} &= \delta_{IJ}
\end{align}
and those of the self-consistent parametrization are
\begin{align}
    A_{IJ}^{[2],\text{sc}} &= \left<\text{HF}\left|\hat{G}_I^\dagger\boldsymbol{U}^\dagger\hat{H}\boldsymbol{U}\hat{G}_J\right|\text{HF}\right>\\\nonumber
    &\quad-\frac{1}{2}\left(\left<\text{HF}\left|\hat{G}_I^\dagger\hat{G}_J\boldsymbol{U}^\dagger\hat{H}\boldsymbol{U}\right|\text{HF}\right> + \left<\text{HF}\left|\boldsymbol{U}^\dagger\hat{H}\boldsymbol{U}\hat{G}_I^\dagger\hat{G}_J\right|\text{HF}\right>\right)\\ B_{IJ}^{[2],\text{sc}} &= \left<\text{HF}\left|\hat{G}_I^\dagger\hat{G}_J^\dagger\boldsymbol{U}^\dagger\hat{H}\boldsymbol{U}\right|\text{HF}\right>\\
    \Sigma_{IJ}^{[2],\text{sc}} &= \delta_{IJ}
\end{align}
In the case of the wave function being an eigenfunction of the Hamiltonian, i.e. $\hat{H}\boldsymbol{U}\left|\text{HF}\right> = E_0\boldsymbol{U}\left|\text{HF}\right>$  or, stated equivalently, when the ansatz parameterization gives the FCI ground-state, $\boldsymbol{U}\left|\text{HF}\right>=\left|\text{FCI}\right>$, the self-consistent working equations reduce to
\begin{align}
    A_{IJ}^{[2],\text{sc*}} &= \left<\text{HF}\left|\hat{G}_I^\dagger\boldsymbol{U}^\dagger\hat{H}\boldsymbol{U}\hat{G}_J\right|\text{HF}\right> -\delta_{IJ}E_0\\
    B_{IJ}^{[2],\text{sc*}} &= 0\\
    \Sigma_{IJ}^{[2],\text{sc*}} &= \delta_{IJ}
\end{align}
with $\text{sc*}$ referring to the assumption that the wave function is an eigenfunction of the Hamiltonian.
It can be seen that, under this assumption, the self-consistent and the state-transfer formulations become identical.

\subsection{LR redundant parameter dependency\label{sec:lr_red}}

In general, the problem solved to find the excitation energies is of the type
\begin{equation}
    \left({\boldsymbol{S}^{[2]}}^{-1}\boldsymbol{E}^{[2]}\right)\boldsymbol{V}_i = \varepsilon_i\boldsymbol{V}_i
\label{eq:LR_geneig}
\end{equation}
This is a
rearrangement of the terms in 
Eq.~(\ref{eq:lr}).
It can be seen that the trace of the left-hand side must be equal to the sum of excitation energies,
\begin{equation}
    \mathrm{tr}\left({\boldsymbol{S}^{[2]}}^{-1}\boldsymbol{E}^{[2]}\right) = \sum_i\varepsilon_i~.
\end{equation}
In the limit of including all possible excitations in the linear response (the FCI limit), this quantity will be constant,
\begin{equation}
    \sum_i^{N_\text{all}}\varepsilon_i = \text{const}~.
\end{equation}
In practice, the linear response equations are truncated early instead of doing the full expansion.
The sum of the excitation energies now becomes dependent on redundant parameters,
\begin{equation}
    \sum_i^{N_\text{truncated}}\varepsilon_i\left(\boldsymbol{\theta}^\text{red},\boldsymbol{\kappa}^\text{red}\right) \neq \text{const}~.
\end{equation}
This implies that the individual excitation energies will depend explicitly on the redundant parameters.

This dependency can be quantified in the case of the state-transfer parameterization, or, if the wave function is an eigenfunction of the Hamiltonian, when using the self-consistent parameterization.
In both cases, the $\boldsymbol{A}^{[2]}$ matrix elements reduce to a unitary transformation of the Hamiltonian shifted by a constant (the ground-state energy)
\begin{equation}
    A^\text{st}_{IJ} = A^\text{sc*}_{IJ} = \left<\text{HF}\left|\hat{G}_I^\dagger\boldsymbol{U}^\dagger_\theta\boldsymbol{U}^\dagger_\kappa\hat{H}\boldsymbol{U}_\kappa\boldsymbol{U}_\theta\hat{G}_J\right|\text{HF}\right> - \delta_{IJ}E_0~,
\label{eq:Ast_Asc}
\end{equation}
where $\boldsymbol{U}_\theta$ is the unitary from the UCC parameterization and $\boldsymbol{U}_\kappa$ is the unitary from the orbital rotation parameterization.
Since the subtraction of the ground-state energy is just a shift of the diagonal, we will ignore this term in the following analysis.
Thus, it can be seen that 
Eq.~(\ref{eq:Ast_Asc}) is identical to performing a unitary transformation of the Hamiltonian,
\begin{equation}
    H_{IJ}\left(\boldsymbol{U}_\kappa,\boldsymbol{U}_\theta\right) = \left<I\left|\boldsymbol{U}^\dagger_\theta\boldsymbol{U}^\dagger_\kappa\hat{H}\boldsymbol{U}_\kappa\boldsymbol{U}_\theta\right|J\right>
\end{equation}
where we introduced the short-hand notation $\left|I\right>=
\hat{G}_I\left|\text{HF}\right>$.
Since a unitary transformation is trace-conserving, we then have
\begin{equation}
    \mathrm{tr}\left\{\boldsymbol{H}\left(\boldsymbol{U}_\kappa,\boldsymbol{U}_\theta\right)\right\} = \sum_{\left|I\right>\in\left|\text{all}\right>}\left<I\left|\boldsymbol{U}^\dagger_\theta\boldsymbol{U}^\dagger_\kappa\hat{H}\boldsymbol{U}_\kappa\boldsymbol{U}_\theta\right|I\right> = \mathrm{const}~. 
\end{equation}
The trace, however, is not conserved for a subspace of the Hamiltonian,
\begin{equation}
\mathrm{tr}\left\{\left[\boldsymbol{H}\left(\boldsymbol{U}_\kappa,\boldsymbol{U}_\theta\right)\right]^\text{SS}\right\} 
= \sum_{\left|I\right>\in\left|\text{SS}\right>}\left<I\left|\boldsymbol{U}^\dagger_\theta\boldsymbol{U}^\dagger_\kappa\hat{H}\boldsymbol{U}_\kappa\boldsymbol{U}_\theta\right|I\right> 
\neq 
\mathrm{const}~.
\label{eq:trace_ss}
\end{equation}
As redundant orbital rotations and redundant cluster amplitudes can perform a unitary transformation of the Hamiltonian, the subspace trace of the Hamiltonian becomes dependent on these redundant parameters.
Since
\begin{equation}
    \sum_i^{N_\text{SS}}\varepsilon_i = \sum_{\left|I\right>\in\left|\text{SS}\right>}\left(\left<I\left|\boldsymbol{U}^\dagger_\theta\boldsymbol{U}^\dagger_\kappa\hat{H}\boldsymbol{U}_\kappa\boldsymbol{U}_\theta\right|I\right> - E_0\right)\label{eq:sum_exc_ss}~,
\end{equation}
\\\noindent
the excitation energies must then also depend on the redundant parameters.

We note that, since the trace of the full Hamiltonian is conserved, if we split our full Hamiltonian into a number of subspaces and sum their traces, then this is equal to the trace of the full Hamiltonian,
\begin{equation}
     \mathrm{tr}\left\{\boldsymbol{H}\left(\boldsymbol{U}_\kappa,\boldsymbol{U}_\theta\right)\right\} = \sum_i^{M_\text{SS}}\mathrm{tr}\left\{\left[\boldsymbol{H}\left(\boldsymbol{U}_\kappa,\boldsymbol{U}_\theta\right)\right]^{\text{SS}_i}\right\} = \text{const}~.
     \label{eq:ss_sum}
\end{equation}
This property does not hold for the naive, projected, and self-consistent parameterizations, which follow the more general case of linear response, as in 
Eq.~(\ref{eq:LR_geneig}).

\subsection{Constrained optimization\label{sec:con_opt}}

From Eq. (\ref{eq:trace_ss}), it can be seen that the ST LR equations can be trace-optimized by doing a minimization of the subspace trace of the Hessian matrix while keeping the ground-state energy constant.
\begin{align}
    \left\{\boldsymbol{\theta}^\text{opt},\boldsymbol{\kappa}^\text{opt}\right\} =& \underset{\boldsymbol{\theta},\boldsymbol{\kappa}}{\mathrm{argmin}}\sum_{\left|I\right>\in\left|\text{SS}\right>}\left<I\left|\boldsymbol{U}^\dagger_\theta\boldsymbol{U}^\dagger_\kappa\hat{H}\boldsymbol{U}_\kappa\boldsymbol{U}_\theta\right|I\right>\label{eq:const_opt}\\
    &\text{subject to}\quad E_0 = \left<\text{HF}\left|\boldsymbol{U}^\dagger_\theta\boldsymbol{U}^\dagger_\kappa\hat{H}\boldsymbol{U}_\kappa\boldsymbol{U}_\theta\right|\text{HF}\right>\nonumber
\end{align}
In the above, the minimization is performed over all $\boldsymbol{\theta}$ and $\boldsymbol{\kappa}$ parameters.
This is done to avoid having to explicitly identify redundant parameters according to 
Eq.~(\ref{eq:kappa_red_def}) and 
Eq.~(\ref{eq:theta_red_def}).
As a practical realization of 
Eq.~(\ref{eq:const_opt}), this minimization problem can be implemented using a penalty function to ensure that the solution found is still the correct ground state wave function, 
\begin{align}
\left\{\boldsymbol{\theta}^\text{opt},\boldsymbol{\kappa}^\text{opt}\right\} &= \underset{\boldsymbol{\theta},\boldsymbol{\kappa}}{\mathrm{argmin}}
\left\{
\sum_{\left|I\right>\in\left|\text{SS}\right>}\left<I\left|\boldsymbol{U}^\dagger_\theta\boldsymbol{U}^\dagger_\kappa\hat{H}\boldsymbol{U}_\kappa\boldsymbol{U}_\theta\right|I\right>\right.\label{eq:penalty_opt}\\\nonumber
&\quad\left.+ K\left(\left<\text{HF}\left|\boldsymbol{U}^\dagger_\theta\boldsymbol{U}^\dagger_\kappa\hat{H}\boldsymbol{U}_\kappa\boldsymbol{U}_\theta\right|\text{HF}\right> - E_0\right)^2
\right\}
\end{align}
with $K$ being a penalty parameter.
The minimization problem can be turned into a maximization by looking for the parameters that satisfy $\mathrm{argmax}$ and changing the sign on the penalty parameter $K$.

\section{Computational Details}

Hartree-Fock calculations to find starting orbitals as well as calculation of molecular integrals were performed using \texttt{PySCF}.\cite{Sun2020-nv}
FCI calculations were performed using \texttt{Dalton}.\cite{Aidas2014-qe}
UCC calculations were performed using  \texttt{SlowQuant}.\cite{slowquant}
All UCC calculations were carried out starting from the Hartree-Fock orbitals, and initial UCC amplitudes were set to zero.
To ensure finding global minima, after convergence, the amplitudes were scaled by a random number between 0 and 10, and the energy was reoptimized again.
This was performed 500 times for each UCC calculation.
All calculations involving penalty functions used a penalty factor of $K=10^{12}$; all calculations where the ground-state energy was not within $10^{-8}$ Hartree of the best found 
ground-state solution were disregarded.
After the penalty function optimization, an energy minimization was performed with respect to the ground state, to ensure that the ground-state wave function was correct.
All UCC optimizations were performed using L-BFGS-B\cite{Zhu1997-ev} through the  \texttt{SciPy}\cite{2020SciPy-NMeth} library.
The basis sets used in the calculations were STO-3G\cite{hehre1969a,hehre1970a} for LiH and 
6-31G\cite{ditchfield1971a} for He.

\section{Results}

In this section, we present an analysis of the different LR parameterizations dependence on the orbital rotations for a FCI wave function of the model system He/6-31G.
This is followed by an analysis of the state-transfer LR parameterization up to singles and doubles for LiH/STO-3G using different levels of truncated UCC expansions.
At last, it proposed to perform a constrained state-averaged optimization to remedy the found problems in truncated LR formulations. 

\subsection{Helium atom}

As a model system to analyze the problem in detail, we first consider the helium atom in the 6-31G basis set, which has one singlet single excitation and one singlet double excitation.
\begin{figure}[H]
    \centering
    \includegraphics[width=0.5\textwidth]{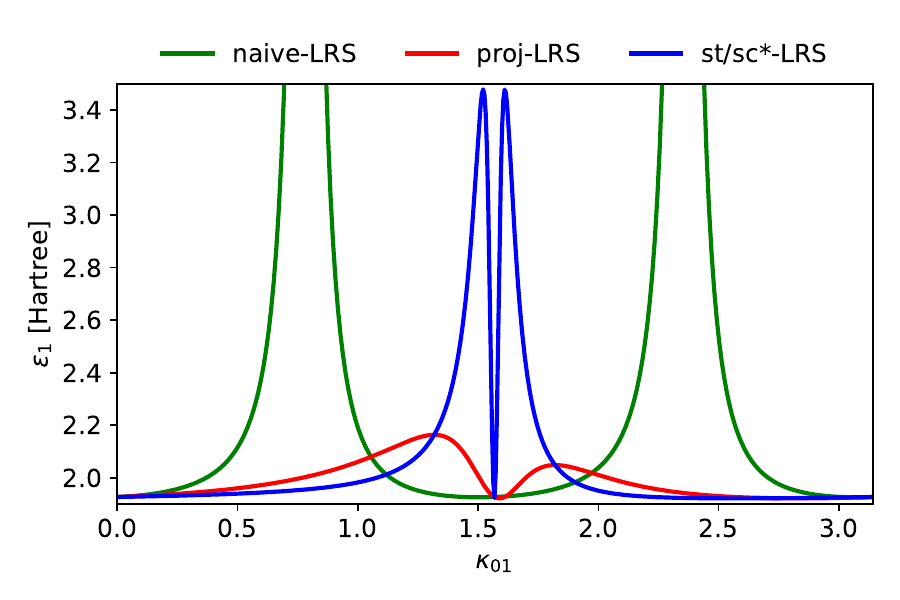}
\caption{Helium. Lowest excitation energy ($\varepsilon_1$), calculated in the 6-31G basis set with LR singlet singles (LRS) and different LR parametrizations, 
    as a function of the redundant orbital rotation parameter $\kappa_{01}$.
    }
    \label{fig:he_lrs}
\end{figure}
Exemplifying the theoretical arguments in section \ref{sec:lr_red}, we show in Fig. \ref{fig:he_lrs} how the singlet single excitation energy depends on the redundant orbital rotation parameter $\kappa_{01}$ for LR truncated to singlet singles. 
The excitation energies obtained with the different LR parametrizations clearly depend on the redundant orbital rotation parameter $\kappa_{01}$.
This shows that all of the truncated LR parameterizations considered here lead to excitation energies that depend on wave function parameters that are redundant with respect to the ground-state energy.
The divergence of the calculated excitation energies using naive-LRS (green line) is due to the metric becoming singular, as shown in previous work.\cite{Kjellgren2024-pc}
\begin{figure}[H]
    \centering
\includegraphics[width=0.5\textwidth]{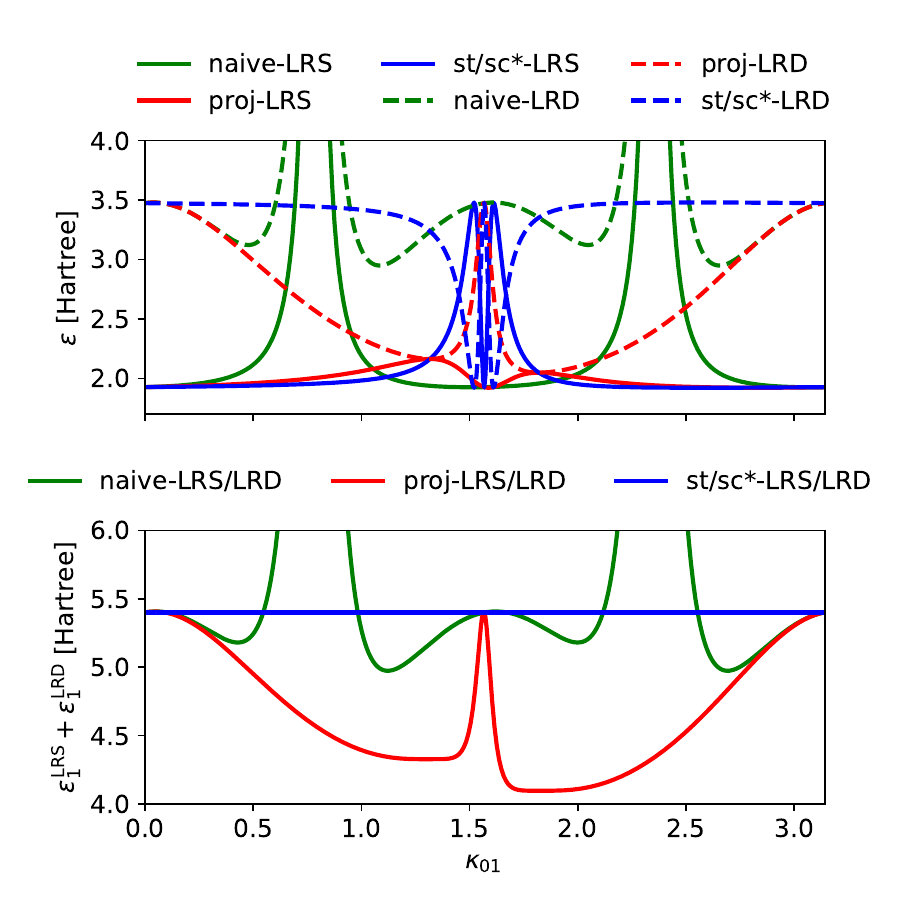}
\caption{The singlet single excitation energy and singlet double excitation energy of the He atom calculated in the 6-31G basis set with LR singlet singles and LR singlet doubles, respectively, as a function of the
redundant orbital rotation parameter $\kappa_{01}$.
The top panel shows the singlet single excitation energy (solid lines), and the singlet double excitation energies (dashed lines).
The bottom panel shows the sum of the singlet single and singlet double excitation energy.}
    \label{fig:he_lrs-lrd}
\end{figure}
To follow our mathematical argument regarding the conservation of subspaces for ST LR in section \ref{sec:lr_red}, we firstly show in Fig.~\ref{fig:he_lrs-lrd} top panel how the excitation energies of He calculated using either LRS (full lines) or LRD (dashed lines) for the different LR parameterizations
vary as a function of the redundant orbital
rotation parameter $\kappa_{01}$.
Next, in the bottom panel, the sum of the LRS and LRD excitation energies is shown.
For the self-consistent$^*$ and state-transfer parameterizations (top panel, blue lines) at around $\kappa_{01}=1.5$, the energies obtained from LRS and LRD flip, that is,
for specific choices of orbitals, the double excitation energy can be calculated using LRS and vice versa.
Looking at the sum of the excitation energies (bottom panel), the two excitation energies add up to a constant value, as described in Eq. (\ref{eq:ss_sum}), for the state-transfer parameterization and self-consistent$^*$ parameterization (blue line).
For the naive (green line) and projected (red line) parameterizations, this is not the case.
This shows that for the st/sc* parameterization, the splitting into subspaces does not affect the sum of all excitation energies.
In contrast, for the naive and projected parameterizations, splitting the Hamiltonian into subspaces does not conserve the sum of excitation energies, due to the metric being different from identity.

\subsection{LiH}

\subsubsection{Trace optimization within LRSD}

Having shown that excitation energies for truncated LR have a dependency on redundant parameters for all of the parameterizations, we will now turn our focus only on the st/sc* parameterization.
Here, we can exploit its property of having the same form as a unitary transformed Hamiltonian, which allows for the constrained optimization outlined in 
Sec.~\ref{sec:con_opt}. 

To showcase that this problem generalizes to systems larger than the helium atom, we construct a set of bad parameters and optimized parameters by performing a constrained optimization of the trace of the unitary transformed Hamiltonian using Eq. (\ref{eq:penalty_opt}).
The bad parameters $\boldsymbol{\theta}^\text{bad}$ and $\boldsymbol{\kappa}^\text{bad}$ are found by doing a maximization, and the optimized parameters $\boldsymbol{\theta}^\text{opt}$ and $\boldsymbol{\kappa}^\text{opt}$ are found by doing a minimization.
\begin{table}[H]
    \centering
    \caption{Trace of the subspace Hamiltonian, $\mathrm{tr}\left\{\left[\boldsymbol{H}\left(\boldsymbol{U}_\kappa,\boldsymbol{U}_\theta\right)\right]^\text{SS}\right\}$, as defined in Eq.~(\ref{eq:trace_ss}), and 
trace of the subspace electronic Hessian, $\mathrm{tr}\left\{\left[\boldsymbol{A}\left(\boldsymbol{U}_\kappa,\boldsymbol{U}_\theta\right)\right]^\text{SS}\right\}$, as defined in Eq.~(\ref{eq:sum_exc_ss}) for LiH in the STO-3G basis set.
In both cases, the subspace is all single and double excited $\hat{S}^2$ conserving states.
All quantities are in Hartree.}
\label{tab:bad_opt_UCC}
    \begin{tabular}{l|c|c|c}
    & oo-UCCSD & oo-UCCSDT & oo-UCCSDTQ \\\hline
$\mathrm{tr}\left\{\boldsymbol{H}^\text{SD}\left(\boldsymbol{\theta}^\text{bad},\boldsymbol{\kappa}^\text{bad}\right)\right\}$ & 
$-$274.60 & $-$274.00 & $-$246.95
\\
$\mathrm{tr}\left\{\boldsymbol{A}^\text{stLRSD}\left(\boldsymbol{\theta}^\text{bad},\boldsymbol{\kappa}^\text{bad}\right)\right\}$ & 
\phantom{$-$}113.93 & \phantom{$-$}114.52 & \phantom{$-$}141.58
\\
$\mathrm{tr}\left\{\boldsymbol{H}^\text{SD}\left(\boldsymbol{\theta}^\text{opt},\boldsymbol{\kappa}^\text{opt}\right)\right\}$ & 
$-$274.60 & $-$274.72 & $-$275.19
\\
$\mathrm{tr}\left\{\boldsymbol{A}^\text{stLRSD}\left(\boldsymbol{\theta}^\text{opt},\boldsymbol{\kappa}^\text{opt}\right)\right\}$ & 
\phantom{$-$}113.93 & \phantom{$-$}113.81 & \phantom{$-$}113.33
\\\hline
\end{tabular}
\end{table}
Table~\ref{tab:bad_opt_UCC} shows the trace of the constrained- minimized and maximized subspace Hamiltonian, $\boldsymbol{H}^\text{SD}$, and subspace Hessian, $\boldsymbol{A}^\text{stLRSD}$.
The subspace contains all possible single and double excited
$\hat{S}^2$-conserving states.
The corresponding sum of excitation energies in this subspace can also be seen as $\mathrm{tr}(\boldsymbol{A}^\text{stLRSD})$.
It is evident that, with increased flexibility in the wave function, the difference between the sum of excitation energies increases when opt-parameter and bad-parameter traces are compared.
For the oo-UCCSD wave function, the difference is $2.27\times10^{-3}$ Hartree; for oo-UCCSDT the difference is 0.71 Hartree, and for oo-UCCSDTQ the difference is 28.25 Hartree.
With a total of 44 excitation energies from the st-LRSD, this corresponds to an average difference in the excitation energies of $0.44$ eV to $17.47$ eV for the oo-UCCSDT and oo-UCCSDTQ, respectively.
As expected, the possibility to either improve or deteriorate the calculated excitation energies increases with the flexibility of the wave function because of increased redundancy with respect to the ground-state energy.
\begin{figure}[H]
    \centering
    \includegraphics[width=0.5\textwidth]{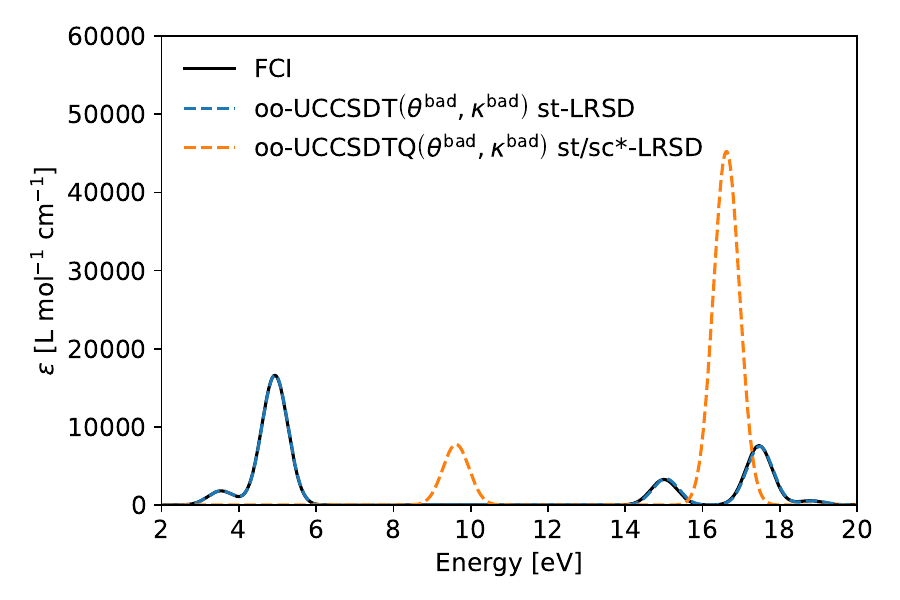}
    \caption{Calculated electronic spectra of LiH in the STO-3G basis using different levels of theory. The oo-UCCSD($\boldsymbol{\theta}^\text{bad},\boldsymbol{\kappa}^\text{bad}$) spectrum is omitted from the figure as it is visually on top of the oo-UCCSDT($\boldsymbol{\theta}^\text{bad},\boldsymbol{\kappa}^\text{bad}$) spectrum.}
    \label{fig:sc_broken}
\end{figure}
To visualize the impact of these energy differences, in Fig. \ref{fig:sc_broken}, the calculated electronic spectra for LiH using st-LRSD with ($\boldsymbol{\theta}^\text{bad},\boldsymbol{\kappa}^\text{bad}$) are shown.
For st-LRSD based on the oo-UCCSDT wave function, even with an average difference of 0.44 eV between the excitation energies obtained using ($\boldsymbol{\theta}^\text{bad},\boldsymbol{\kappa}^\text{bad}$) and those obtained using ($\boldsymbol{\theta}^\text{opt},\boldsymbol{\kappa}^\text{opt}$), there is no visual difference between the ($\boldsymbol{\theta}^\text{bad},\boldsymbol{\kappa}^\text{bad}$) spectra (blue dashed line) and the FCI spectra (black line) at low excitation energies.
This is because the difference is dominated by higher-lying excitations, in the 60 eV to 180 eV range.
For st-LRSD based on the oo-UCCSDTQ, the difference between ($\boldsymbol{\theta}^\text{bad},\boldsymbol{\kappa}^\text{bad}$) and ($\boldsymbol{\theta}^\text{opt},\boldsymbol{\kappa}^\text{opt}$) is also dominated by the higher-lying excitations; however, due to the increased redundancy of the parameters in the wave function, even the lowest-lying excitations (orange dashed line) are significantly different from the FCI references (black line).

The possibility of getting erroneous excitation energies even when the ground-state solution is the FCI wave function could make schemes that take advantage of redundant orbital rotations, such as orbital localization,\cite{Sayfutyarova2017-zi,Aquilante2006-oi,Lowdin1950-nx,Sun2014-hm,Reed1985-ur,Knizia2013-jw,Foster1960-rd,Pipek1989-gz,Edmiston1963-aq} difficult to combine with truncated LR.
In recent work by Grimsley and Evangelista,\cite{Grimsley2024-qo} for instance, it was shown that using small expansions with sc-qEOM was sometimes unstable when combined with an ADAPT wave function; this could be caused by the ambiguity of the redundant wave function parameters with respect to the ground-state energy, as shown in this work.  

\subsubsection{Trace optimization within LRS}

In the previous section, the low-energy spectrum with opt-parameters improved only marginally compared to the spectrum obtained from the bad-parameters for both the oo-UCCSD and oo-UCCSDTQ wave functions.
Here, we want to focus and understand how the optimization procedure can positively impact the low energy spectrum compared to HF orbitals, as HF orbitals are often the orbitals used as a basis for correlated calculations.
To highlight this, we switch to a smaller LR expansion of only single excitations.
\begin{figure}[H]
    \centering  \includegraphics[width=0.5\textwidth]{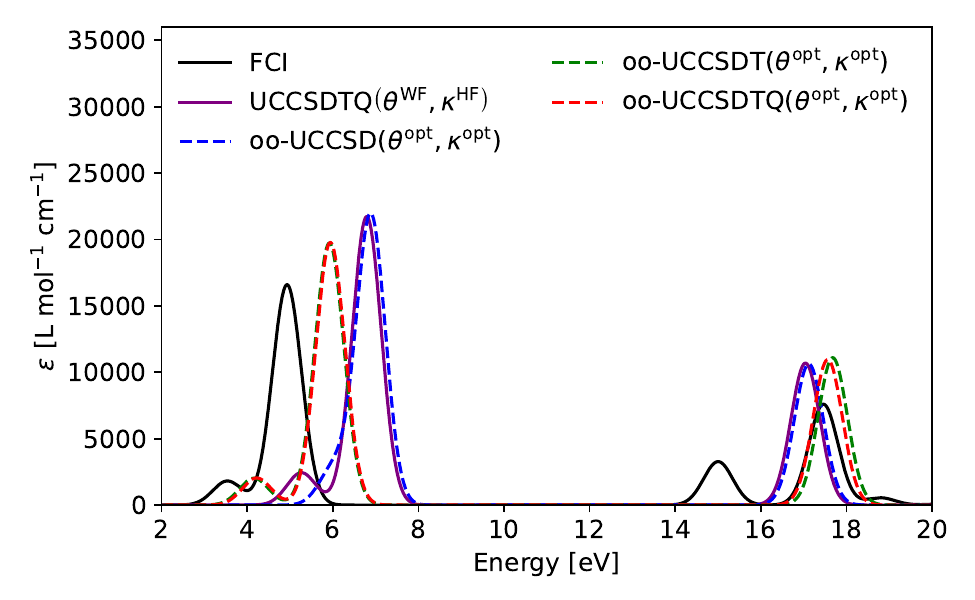}
    \caption{LiH/STO-3G.  Spectra calculated with state-transfer LR singles (st-LRS) using various UCC wave functions.
$\boldsymbol{\theta}^\textrm{WF}$ refers to the $\boldsymbol{\theta}$-values found by only optimizing for the ground-state.}
    \label{fig:stlrs}
\end{figure}
In Fig. \ref{fig:stlrs}, the calculated spectra of LiH using st-LRS for different UCC expansions can be seen.
Comparing the spectra calculated using st-LRS based on the oo-UCCSD($\boldsymbol{\theta}^\text{opt},\boldsymbol{\kappa}^\text{opt}$) wave function (blue dashed-line) to those calculated using st-LRSD based on the UCCSDTQ ($\boldsymbol{\theta}^\text{WF},\boldsymbol{\kappa}^\text{HF}$) wave function (purple line) shows that the lack of flexibility in the oo-UCCSD wave function makes it constrained to be close to the solution found using Hartree-Fock orbitals with an UCCSDTQ expansion.
Comparing the spectra calculated using st-LRS based on oo-UCCSDT($\boldsymbol{\theta}^\text{opt},\boldsymbol{\kappa}^\text{opt}$) (green dashed-line) and on oo-UCCSDTQ($\boldsymbol{\theta}^\text{opt},\boldsymbol{\kappa}^\text{opt}$) (red dashed-line) with the st-LRS spectra based on the UCCSDTQ($\boldsymbol{\theta}^\text{WF},\boldsymbol{\kappa}^\text{HF}$) wave function (purple line), it can be seen that the first peak around 4-8 eV is improved by 0.85 eV, and the third peak around 16-18 eV is improved by 0.32 eV (the error of the UCCSDTQ to FCI is $-$0.42 eV and the error of the oo-UCCSDTQ to FCI is 0.10 eV).
It should be noted that the peak around 15 eV is missing from all st-LRS calculations.
These results highlight that the Hartree-Fock orbitals are not always near-optimal orbitals to use together with a truncated LR model.
The discrepancy to the FCI spectra (black line) is due to insufficient flexibility in the wave function and/or LR expansion.
That is, the spectra can be improved by doing a larger LR expansion as seen in the st-LRSD results, but they can also be improved by adding more parameters to the wave function that are redundant with respect to the ground-state energy, that is, by doing generalized UCC.
\begin{table}[H]
    \centering
\caption{Trace of the subspace Hamiltonian as defined in Eq.~(\ref{eq:trace_ss}), and trace of the subspace electronic Hessian as defined in Eq.(\ref{eq:sum_exc_ss}) for LiH in the STO-3G basis set. For both, the subspace is all single excited $\hat{S}^2$ conserving states.
All quantities are in Hartree.}\label{tab:opt_lrs}
\begin{tabular}{l|c|c|c}
    & oo-UCCSD & oo-UCCSDT & oo-UCCSDTQ
\\\hline
$\mathrm{tr}\left\{\boldsymbol{H}^\text{S}\left(\boldsymbol{\theta}^\text{opt},\boldsymbol{\kappa}^\text{opt}\right)\right\}$ & $-$60.38 & $-$60.48 & $-$60.48
\\
$\mathrm{tr}\left\{\boldsymbol{A}^\text{stLRS}\left(\boldsymbol{\theta}^\text{opt},\boldsymbol{\kappa}^\text{opt}\right)\right\}$ & \phantom{$-$}10.26 & \phantom{$-$}10.16 & \phantom{$-$}10.16
\\\hline
    \end{tabular}
\end{table}
In Table~\ref{tab:opt_lrs} we report the constrained minimized trace of the subspace Hamiltonian where the subspace contains all possible single excited $\hat{S}^2$ conserving states.
Compared to oo-UCCSD, the increased flexibility of oo-UCCSDT/oo-UCCSDTQ on average lowers the excitation energies by 0.34 eV (8 st-LRS excitations in total).
Clearly, the oo-UCCSDT and oo-UCCSDTQ $(\boldsymbol{\theta}^\text{opt},\boldsymbol{\kappa}^\text{opt})$ excitation energies sum to the same value (within four digits), and this could be caused by the quadruple excitations being unable to directly couple to the single excited determinants in the wave function. That is, quadruple excitations can only improve the single excited block indirectly through the improvement of the double and triple excited determinants.

\subsubsection{CSA-UCC}

Instead of minimizing the trace of the unitary transformed Hamiltonian with respect to all states included in the LR expansion, the minimization can be performed only for a few selected states, which corresponds to restricting $\left|\text{SS}\right>$ to only be some specific states in 
Eq.~(\ref{eq:penalty_opt}).
For this example, the focus will be on the states corresponding to the first two peaks.
The states to minimize are identified as the three smallest diagonal elements in $\boldsymbol{A}^\text{stLRS}(\boldsymbol{\theta}^\text{WF},\boldsymbol{\kappa}^\text{HF})$, and correspond to the CSFs $\frac{1}{\sqrt{2}}\left(\left|1110 0100 0000\right>-\left|1101 1000 0000\right>\right)$, $\frac{1}{\sqrt{2}}\left(\left|1110 0001 0000\right>-\left|1101 0010 0000\right>\right)$, and, $\frac{1}{\sqrt{2}}\left(\left|1110 0000 0100\right>-\left|1101 0000 1000\right>\right)$.
This procedure can be seen as a constrained state-averaged UCC (CSA-UCC), with the name derived from state-averaged VQE\cite{Parrish2019-wf,Nakanishi2019-qf,Grimsley2025-su,yalouz2021state} (SA-VQE) in the context of circuit based wave functions.
\begin{figure}[H]
    \centering
    \includegraphics[width=0.5\textwidth]{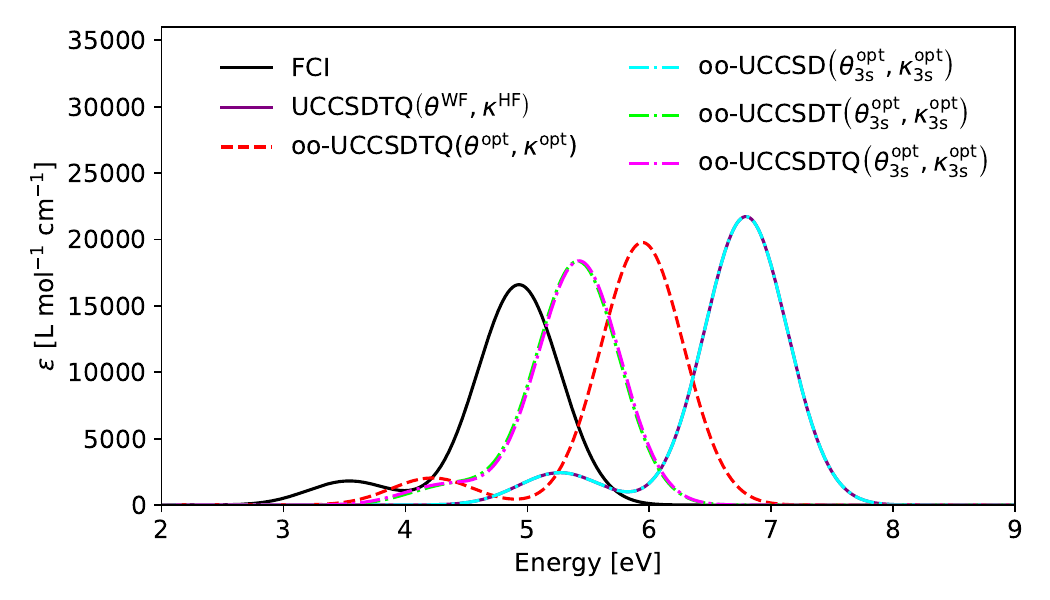}
    \caption{LiH/STO-3G. Spectra calculated with state-transfer LR singles (st-LRS) using various UCC wave functions, where the wave function parameters are constrained optimized for the first three excited states denoted with $(\boldsymbol{\theta}^\text{opt}_\text{3s}, \boldsymbol{\kappa}^\text{opt}_\text{3s})$.
    $\boldsymbol{\theta}^\text{WF}$ refers to the $\theta$-values found by only optimizing for the ground-state, this is the same spectra as UCCSDTQ($\boldsymbol{\theta}^\text{WF},\boldsymbol{\kappa}^\text{HF}$) in Fig. \ref{fig:stlrs}.
    $\boldsymbol{\theta}^{\text{opt}}$ and $\boldsymbol{\kappa}^{\text{opt}}$ are parameters optimized for all 8 excitations in LRS, i.e. the spectra as can be seen in Fig. \ref{fig:stlrs}.
    }
    \label{fig:stlrs_3states}
\end{figure}
In Fig. \ref{fig:stlrs_3states}, the calculated LiH spectra can be seen for different UCC expansions and different optimizations.
When the constrained minimization of the trace of the subspace Hamiltonian is applied only to the three states corresponding to the first two peaks, it can be seen that the calculated spectra move closer to that of the FCI solution, than the spectra found when doing the constrained minimization of the trace of the singlet singles and single doubles subspace Hamiltonian.
The error in the tallest peak is 1.87 eV, 1.02 eV, and 0.50 eV with respect to FCI for the UCCSDTQ($\boldsymbol{\theta}^\text{WF},\boldsymbol{\kappa}^\text{HF}$), oo-UCCSDTQ($\boldsymbol{\theta}^{\text{opt}},\boldsymbol{\kappa}^{\text{opt}}$) and oo-UCCSDTQ($\boldsymbol{\theta}^\text{opt}_\text{3s},\boldsymbol{\kappa}^\text{opt}_\text{3s}$), respectively.
This improvement shows that LR expansions can possibly be truncated early if the wave function is optimized to capture the relevant excitations.

In this work, the focus was on the UCC wave function.
However, the findings also apply to quantum circuit-based wave functions, as these are also unitary by construction.
Using a unitary product state wave function, such as those representable by a quantum circuit, can give more flexibility to the wave function parameterization than the UCC expansion and might allow for accurate excitation spectra using only a small LR expansion.
Performing constrained SA-VQE (CSA-VQE) in combination with LR to optimize the use of quantum resources will be part of future research.

\section{Conclusion}

In this work, we show that the excitation energies calculated using a truncated LR formalism have an explicit dependency on redundant parameters (with respect to the ground-state energy) in a unitary parameterized wave function.
This dependency was shown numerically  to be present for four different LR parameterizations,
namely naive, projected, self-consistent, and state-transfer,
by considering a minimal system (He in the 6-31G basis set).

For the state-transfer and self-consistent (in the limit of the wave function being an eigenfunction of the electronic Hamiltonian) parametrizations, it was shown that the LR equations take the form of a shifted unitary transformed Hamiltonian matrix.
This particular form of the LR equations was used to set up a constrained minimization/maximization problem that relies only on the diagonal elements of the LR matrix.
This formalism was used to show the problem of dependency on redundant parameters generalized to larger systems (LiH in the STO-3G basis set) when using truncated LR, here truncated to singles and doubles.
Finally, it was shown that the constrained minimization formalism through CSA-UCC can improve the results of smaller LR expansions.
Combining CSA with VQE might lead to better use of quantum resources than a pure LR approach, and would require further research.

\acknowledgments
We acknowledge the support of the Novo Nordisk Foundation (NNF) for the focused research project ``Hybrid Quantum Chemistry on Hybrid Quantum Computers'' (grant number  NNFSA220080996).

\section*{DATA AVAILABILITY}
The data that support the findings of this study are available from the corresponding author upon reasonable request.

\newpage
\bibliographystyle{unsrt}
\bibliography{literature}

\end{document}